# Comprehensive Analysis of Thermal Dissipation in Lithium-Ion Battery Packs

Xuguang Zhang[a], Hexiang Zhang[a], Amjad Almansour[b], Mrityunjay Singh[c], Hengling Zhu[a], Michael C. Halbig[b,*], Yi Zheng[a,d,**]

[a] *Department of Mechanical and Industrial Engineering, Northeastern University, Boston, MA 02115, USA*

[b] *NASA Glenn Research Center, Cleveland, OH 44135, USA*

[c] *Ohio Aerospace Institute, Cleveland, OH 44142, USA*

[d] *Department of Chemical Engineering, Northeastern University, Boston, MA 02115, USA*

[*]**E-mail:** michael.c.halbig@nasa.gov

[**]**E-mail:** y.zheng@northeastern.edu

## ABSTRACT

Effective thermal management is critical for lithium-ion battery packs' safe and efficient operations, particularly in applications such as drones, where compact designs and varying airflow conditions present unique challenges. This study investigates the thermal performance of a 16-cell lithium-ion battery pack by optimizing cooling airflow configurations and integrating phase change materials (PCMs) for enhanced heat dissipation. Seven geometric configurations were evaluated under airflow speeds ranging from 0 to 15 m/s, reflecting the operational conditions of civilian drones. A comprehensive 3D simulation approach was used to analyze the effects of inlet and outlet configurations, airflow dynamics, and PCM phase transition behavior. Results indicate that the trapezoidal (wide-base) configuration, paired with a 5-inlet and 1-outlet setup, achieves the most balanced performance, effectively maintaining optimal operating temperatures across low and high-speed airflow conditions. PCM integration further stabilized thermal behavior, with phase change durations extending to 12.5 min under tested conditions. These findings highlight the importance of geometric optimization and material integration in advancing compact and reliable thermal management systems for energy-dense battery packs. This study provides a foundation for designing efficient cooling strategies tailored to lightweight applications such as drones and portable energy storage systems.

## 1. Introduction

The increasing demand for energy-dense lithium-ion battery systems in applications such as electric vehicles (EVs), drones, and renewable energy storage highlights the critical need for advanced thermal management solutions.[1,2] Lithium-ion batteries, while offering high energy efficiency and long cycle life, are particularly vulnerable to thermal fluctuations, which can reduce performance, shorten lifespan, and lead to safety risks such as thermal runaway.[3,4] Effective heat dissipation during charging and discharging cycles is essential for ensuring safe and reliable operation in compact battery configurations.[5,6]

Among various cooling techniques, forced air cooling and phase change material (PCM)-based strategies have emerged as effective solutions.[7,8] Due to its simplicity and cost-efficiency, forced air cooling is desirable for lightweight applications such as drones and portable devices.[9] However, achieving uniform airflow and minimizing hotspots in densely packed configurations remains a significant challenge.[10,11] For maintaining optimal battery performance and safety, PCM integration has been shown to effectively absorb excess heat during phase transitions, stabilizing battery temperatures during peak loads, though its efficiency depends heavily on the material's thermal properties and phase change dynamics.[12,13,14] As the use of drones increases in daily life, the requirements for heat dissipation and battery life are becoming more stringent. Civilian drones typically operate within a speed range of 0 to 15 m/s, depending on their designs and applications.[15,16] The airflow speed range in this study is controlled based on these operational characteristics to ensure relevance to real-world scenarios.

Recent research has explored hierarchical and micro-structured materials for sustainable cooling applications. For instance, Shi et al. demonstrated the potential of hierarchically micro- and nanostructured polymers in enhancing passive cooling, paving the way for lightweight and scalable thermal solutions.[17] Like those developed by Li et al., transparent aerogels offer novel opportunities for integrating lightweight, thermally efficient materials in energy systems.[18,19] These advancements underscore the importance of combining material innovations with optimized thermal management designs.

This study presents a novel approach to thermal management for a 16-cell lithium-ion battery pack, leveraging systematic optimization of airflow configurations and PCM integration to enhance cooling performance. Building on the prior work highlighting the effectiveness of hybrid heat dissipation systems,[20] this research investigates the synergistic effects of airflow geometry, PCM phase transition dynamics, and battery pack configurations under low-speed airflow conditions relevant to drone operations.[21,22] Distinguishing from earlier studies that emphasize individual techniques, this work utilizes the simulation software Ansys Discovery 2024 R1 to comprehensively evaluate seven distinct geometric configurations under airflow speeds ranging from 0 to 15 m/s.[23,24]

The innovation of this study lies in the detailed analysis of airflow dynamics and geometric optimization for battery arrangements. Among the configurations examined, the trapezoidal (wide-base) configuration was identified as the most effective in balancing thermal performance across low- and high-speed regions. Additionally, the integration of PCM with enhanced thermal conductivity and latent heat properties, as studied in the prior research,[25] prolongs the phase change duration, improving the heat dissipation during extended operation.[26] By addressing the interplay among material properties, geometric optimization, and airflow design, this study advances state-of-the-art thermal management systems for lithium-ion battery packs, enabling safer and more efficient energy storage solutions.[27,28,29,30]

## 2. Design and simulation

The analysis steps are presented in Fig. 1(a), outlining the progression from initial tests to comprehensive simulations. The study begins by evaluating a single 18650 battery capsule, analyzed through simulation and experimental methods to validate the computational approach.[8] This research centers on the thermal performance of a 16-battery pack, where each battery capsule is modeled as a uniform cylindrical structure. Key features explored include variations in battery arrangements, the influence of airflow inlet and outlet numbers, and the effect of airflow speed on heat dissipation. Simulations were conducted using Ansys Discovery 2024 R1 and Ansys Workbench 2024 R1.[31] Determining optimal inlet and outlet numbers was a preliminary step to ensure adequate airflow and thermal management before finalizing the 16-battery pack design. The chosen arrangement, the trapezoidal (wide-base) configuration, was selected for its superior thermal performance.

Fig. 1(b) provides a detailed schematic of the trapezoidal (wide-base) configuration arrangement, highlighting its structural design. The 3D-printed shell encloses the surrounding surfaces of the battery pack, ensuring a secure airflow area while enhancing the structural integrity. Battery positions within the arrangement are evenly distributed to ensure uniform heat transfer and airflow paths.

Simulation results are presented in Fig. 1(c) and Fig. 1(d), showing the temperature distribution across the 16-battery pack. Airflow direction remains consistent in both figures to facilitate comparative analysis. Fig. 1(c) provides an isometric view of the 3D simulation results, while Fig. 1(d) presents a top view of the same simulation. These visualizations are included for display purposes, with all results analyzed comprehensively in the simulation results section.

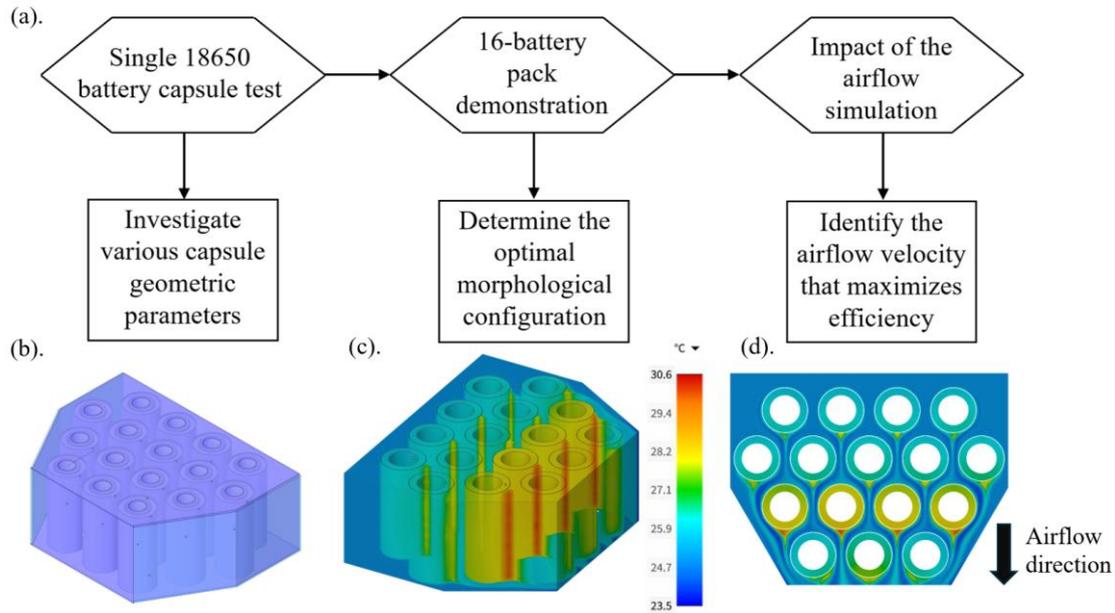

**Fig. 1** (a). Research steps. (b). Isotropic view of the trapezoidal (wide-base) configuration. (c). Isotropic view of the simulation result of the trapezoidal (wide-base) configuration. (d). Top view of the trapezoidal (wide-base) configuration.

## 2.1. Base parameters setting

The diamond shape is an ideal symmetric configuration for this study, providing uniform airflow distribution and balanced geometric characteristics for thermal analysis. In Fig. 2, 5 inlets are positioned symmetrically along the top of the battery pack, and 5 outlets are located at corresponding points along the bottom. The initial diamond shape configuration involved a single inlet (Inlet 1) paired with a single outlet (Outlet 1). Subsequently, to investigate the influence of inlet location on thermal performance, the outlet was fixed at Outlet 1 while the inlet position was sequentially varied from Inlet 1 to Inlet 5. This process was reversed to examine the effect of outlet location, with the inlet fixed at Inlet 1 and the outlet position varied from Outlet 1 to Outlet 5. These sequential tests provided insights into the most effective inlet-outlet pairings for maximizing airflow efficiency and thermal dissipation.

Additionally, Fig. 2(b) highlights the structural details of the battery sheath and the materials applied in the simulation. The sheath design is defined by 4 key radii: the inner sheath inside radius ($r_{is,i}$), the inner sheath outside radius ($r_{is,o}$), the outer sheath inside radius ($r_{os,i}$), and the outer sheath outside radius ($r_{os,o}$). The layers of the sheath are color-coded to represent their respective materials: the blue region corresponds to Markforged's Onyx$^©$ filament, a composite material reinforced with chopped carbon fibers. This material is known for its exceptional mechanical strength, high stiffness, and compatibility with 3D printing technologies, making it an ideal choice for securing the structural integrity of the battery pack. The red region denotes the nano-carbon-based PCM, selected for its superior thermal conductivity and latent heat storage properties, which are critical for effective thermal management.

This configuration ensures structural integrity and optimizes the airflow channel and thermal management performance. This study establishes a foundation for achieving a high-efficiency heat dissipation system in battery packs by combining a systematic analysis of inlet-outlet positioning and advanced material integration.

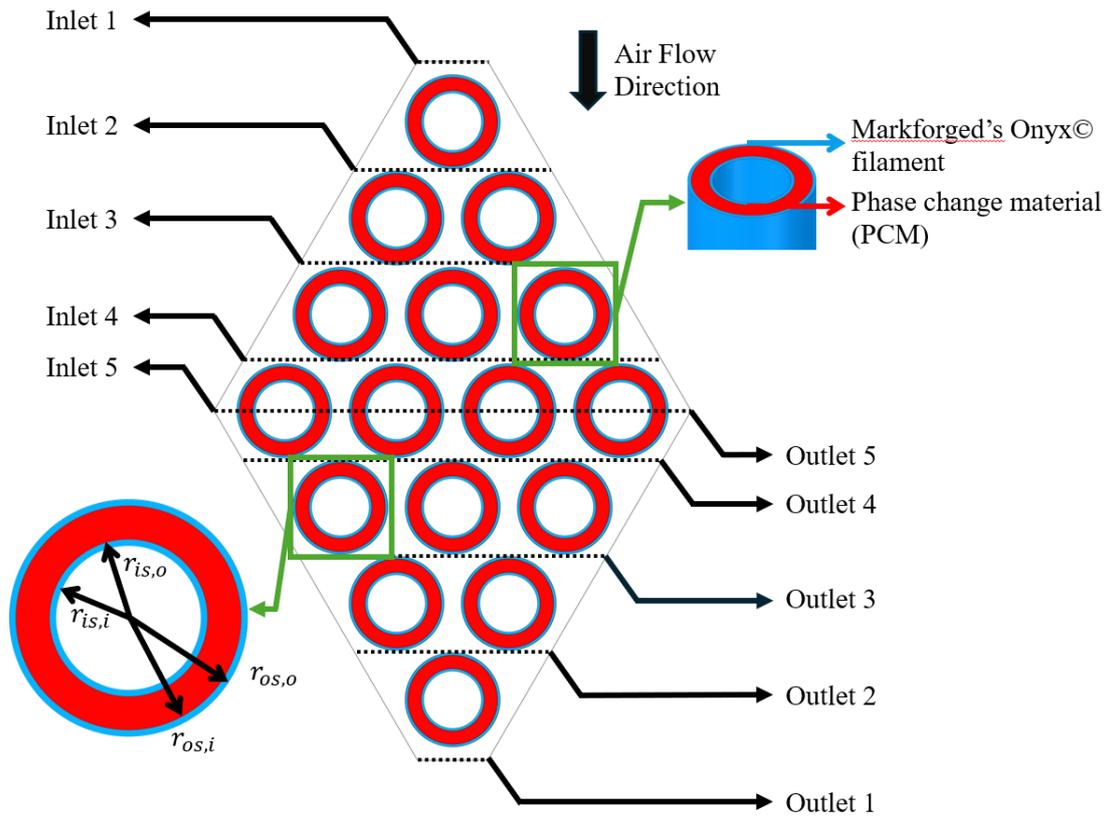

**Fig. 2** Schematic representation of the diamond-shaped configuration with inlet and outlet arrangements, battery sheath radius definitions, and material specifications used in the simulation.

## 2.2. The 16-battery pack configuration

For a 16-battery pack, the initial configuration considered is the 4×4 square arrangement (4444, Fig. 3(h)). This configuration is one of the most common and straightforward designs due to its simplicity in alignment and space efficiency. All battery units in this setup are arranged in uniform rows and columns, providing a baseline for comparison with other configurations. The irregular rectangular configuration (4444-IR, Fig. 3(c)) is generated by transforming this configuration and introducing a horizontal offset to each row. This staggered arrangement improves airflow distribution around each battery unit. Rotating the irregular rectangular configuration 90 degrees clockwise produces the inverted irregular rectangular configuration (4444-IIR, Fig. 3(e)), which retains the staggered nature while providing an alternative flow pattern.

The diamond configuration (1234321, Fig. 3(a)) is another critical design considered in this study. Its symmetrical shape is ideal for investigations requiring uniform geometric characteristics, making it valuable for symmetry-focused thermal and airflow research. This configuration allows for balanced heat dissipation and airflow paths, offering insights into performance under symmetric boundary conditions.

To achieve optimal cooling, staggered arrangements are prioritized. Staggering ensures airflow can reach all battery units more effectively, reducing thermal hotspots. It is well-known that heat tends to accumulate in the tail region of airflow setups as the air temperature rises while absorbing heat from upstream battery units. By addressing this, configurations with narrow tails were explored to increase airflow velocity in the tail region, which aids in mitigating heat accumulation.

The trapezoidal (wide-base) configuration (4543, Fig. 3(f)) introduces a staggered arrangement with a broader base and a narrower tail. This configuration balances airflow distribution while leveraging a funnel-like effect at the tail to increase airflow speed, thus enhancing heat dissipation in the downstream region. It is also efficient in space utilization with its compact four-row design. Transforming this design to a trapezoidal (narrow-top) configuration (43432, Fig. 3(d)) creates an even narrower tail, further accelerating airflow velocity at the downstream end and improving cooling performance.

An alternative transformation of the wide-base configuration is the trapezoidal (narrow-mid) configuration (5434, Fig. 3(b)). This arrangement creates a funnel shape, maintaining the narrow tail while slightly increasing the width at the midpoint. This design compromises airflow acceleration and uniform cooling throughout the pack.

Lastly, the funnel configuration (54322, Fig. 3(g)) maximizes the funnel effect with a significantly narrow tail region that achieves the highest airflow velocity among the configurations. This setup is particularly effective in preventing thermal accumulation at the tail but sacrifices space efficiency due to its less compact design.

Each configuration is selected and tested to evaluate its impact on the thermal heat dissipation performance. The variety of the shapes allows for a comprehensive analysis of how geometric and staggered arrangements influence the cooling efficiency of a 16-battery pack under uniform operating conditions.

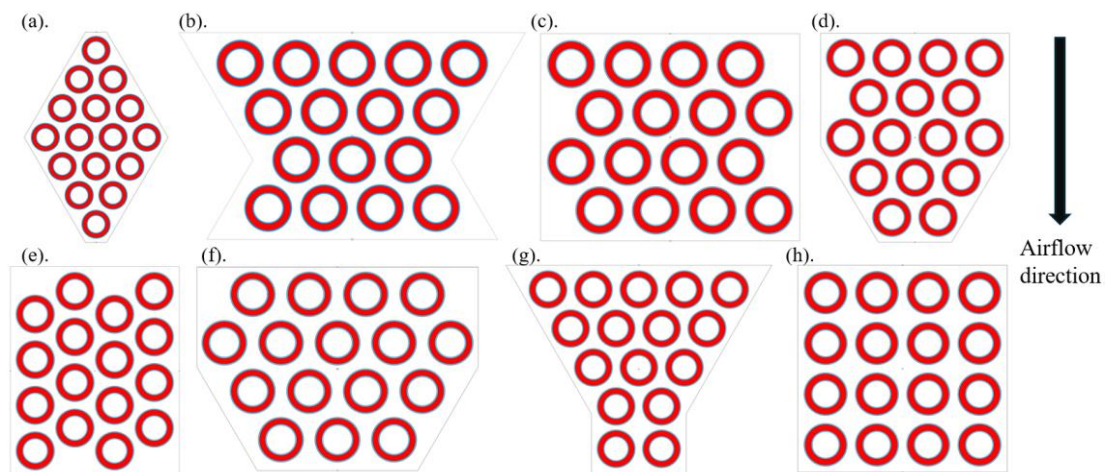

**Fig. 3** (a). Diamond configuration (1234321). (b). Trapezoidal (Narrow-Mid) configuration (5434). (c). Irregular Rectangular Configuration (4444-IR). (d). Trapezoidal (Narrow-Top) Configuration (43432). (e). Inverted Irregular Rectangular Configuration (4444-IIR). (f). Trapezoidal (Wide-Base) Configuration (4543). (g). Funnel Configuration (54322). (h). Rectangular Configuration (4444).

### 2.3. Ansys thermal simulation settings

A mesh independence study was performed to ensure that the simulation results were not significantly influenced by the mesh density, thereby validating the accuracy and computational efficiency of the numerical model. The study analyzed the effect of mesh unit size on key thermal and airflow performance indicators, such as maximum surface temperature, temperature gradients, and airflow velocity profiles. The initial coarse mesh had a unit size of 0.005 m, resulting in 37,003 mesh units. Subsequently, the mesh density was progressively refined to the finest mesh, with a unit size of

0.0005 m and 9,268,064 mesh units. A series of simulations were conducted across this range to identify the point at which further refinement produced negligible changes in the results. The final mesh chosen for the study had a unit size of 0.001 m, comprising 1,135,622 mesh units. This mesh density was selected to balance computational efficiency and result accuracy optimally. The differences in key performance metrics, the maximum temperature, between this configuration and the finest mesh were within 3%, which is considered negligible for this study. Fig. 4 illustrates three computational mesh samples, highlighting the refinement of near-critical regions where steep gradients occur, including battery sheath surfaces, PCM layers, and airflow channels. The selected mesh ensures accurate thermal and airflow characteristics resolution while maintaining a manageable computational cost. This approach provides the reliability and robustness of the simulation results presented in this study, confirming that the outcomes are independent of the mesh resolution.

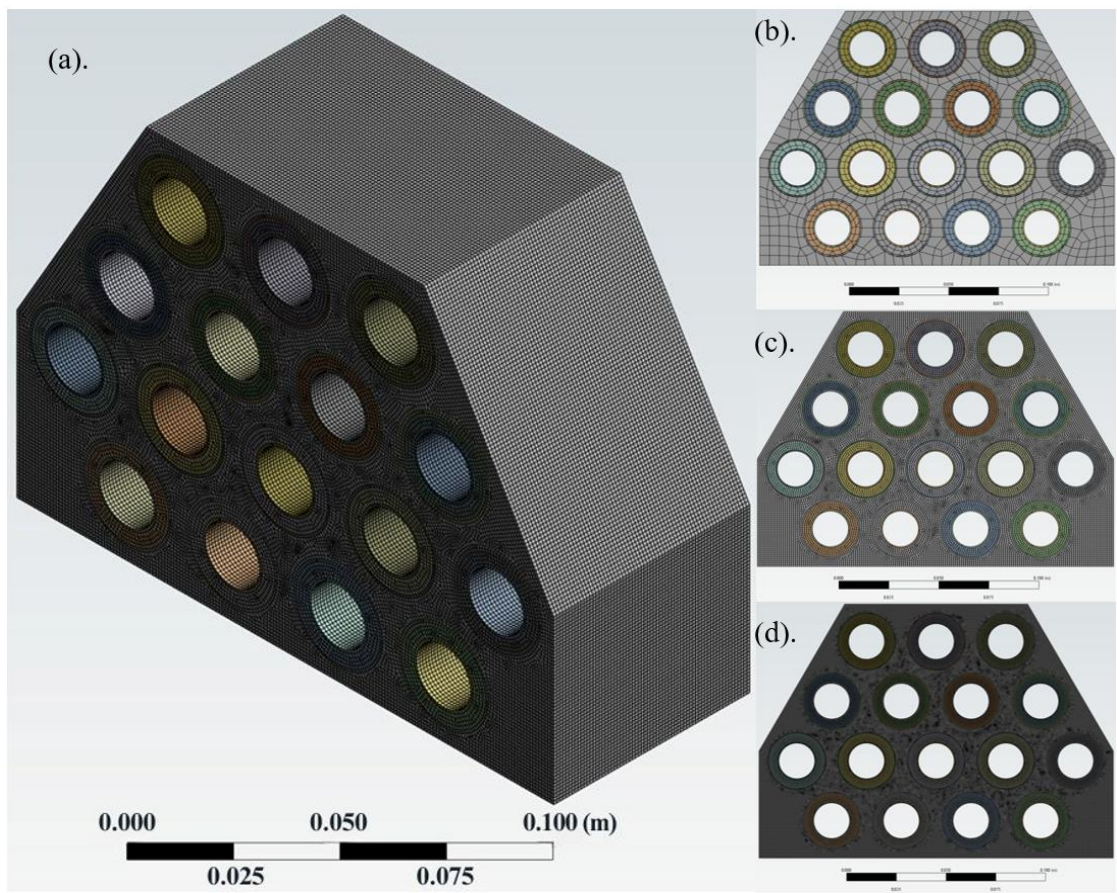

**Fig. 4** (a). Isotropic view of the mesh resolution size of 0.001 m. (b). Top view of the mesh resolution size of 0.005 m. (c). Top view of the mesh resolution size of 0.001 m.

(d). Top view of the mesh resolution size of 0.0005 m. All figures share the same scale bar.

The simulation parameters listed in Table 1 define the thermal and physical properties used for the 16-battery pack analysis. The operating temperature range of the batteries is set between 10 °C and 55 °C, with an optimal operating temperature of 45 °C, ensuring the conditions align with practical operational limits. A uniform surface heat flux of 1,322.88 W/m² is applied to represent the heat generation during operation, with an initial temperature of 25 °C and an inlet airflow velocity varies from 0 to 15 m/s.

For the PCM, the melting temperature is set at 40°C, with a latent heat of 173,400 J/kg, specific heat of 2,890 J/kg · K, density of 800 kg/m³, and thermal conductivity of 16.6 W/m · K. These properties ensure the PCM's capacity to absorb and regulate heat effectively. The sheath material is characterized by a thermal conductivity of 0.13 W/m · K, density of 1,380 kg/m³, and specific heat of 1,420 J/kg · K. These values reflect the thermal insulation and structural stability required to contain and manage heat transfer in the system.

Table 1: Thermal properties and operating setting of the simulation

| Parameters | Value | Units |
|---|---|---|
| Battery operating temperature range | 10 - 55 | °C |
| Optimal operating temperature $T_{ideal}$ | 45 | °C |
| Surface heat flux $q_{gen}$ | 1,322.88 | W/m² |
| Initial temperature $T_i$ | 25 | °C |
| Inlet velocity V | 0-15 | m/s |
| PCM melting temperature $T_{PCM}$ | 40 | °C |
| PCM latent heat $\Delta H$ | 173,400 | J/kg |
| PCM specific heat $c_p$ | 2,890 | J/kg·K |
| PCM density $\rho_{PCM}$ | 800 | kg/m³ |
| PCM thermal conductivity $k_{PCM}$ | 16.6 | W/m·K |
| Sheath thermal conductivity $k_s$ | 0.13 | W/m·K |
| Sheath density $\rho_s$ | 1380 | kg/m³ |
| Sheath specific heat $c_{p_s}$ | 1420 | J/kg·K |
| Inner sheath inside radius $r_{is,i}$ | 18 | mm |
| Inner sheath outside radius $r_{is,o}$ | 19.8 | mm |
| Outer sheath inside radius $r_{os,i}$ | 28 | mm |
| Outer sheath outside radius $r_{os,o}$ | 29.8 | mm |

## 3. Simulation results

The simulation results focus on evaluating the cooling performance of a 16-battery pack under different configurations and airflow conditions, particularly within the low, mid, and high-speed airflow regions. This range is critical for drone operations, especially during takeoff and hovering, where effective thermal management is essential. The study investigates the influence of the inlet and outlet configurations, followed by a comparative analysis of seven distinct battery pack arrangement configurations. All maximum temperature values were obtained under steady-state conditions, with the highest temperature consistently located at the tail unit of the battery pack across all configurations. This finding underscores the significance of addressing thermal accumulation at the downstream end of the airflow path.

The analysis begins with the determination of the optimal number of inlets. In these simulations, the outlet was fixed at a single location (one outlet) while varying the number and positions of the inlets, as illustrated in Fig. 2. The data in Fig. 5(a) indicate that one inlet results in a slow heat dissipation effect, leading to higher maximum temperatures. As the number of inlets increases, the cooling effect improves due to the enhanced airflow distribution. However, the configurations with 4 or 5 inlets show diminishing returns in the high-speed region, likely due to turbulence and localized hotspots. In the low-speed region, the configuration with 5 inlets consistently demonstrated the best heat dissipation performance, maintaining the lower maximum temperatures compared to other setups. Based on these findings, the 5-inlet configuration was adopted for the subsequent analyses.

Following the inlet analysis, the number of outlets was varied to determine its impact on the cooling performance. In these simulations, the inlet configuration was fixed as one inlet. As shown in Fig. 5(b), reducing the number of outlets improved the cooling effect. The one-outlet configuration, in particular, exhibited slightly better cooling performance than the two-outlet setup. This result can be attributed to the funnel-like airflow pattern generated by the narrow tail in the one-outlet design, accelerating the airflow and enhancing heat dissipation. As the number of outlets increases, the laminar airflow effect decreases, while the turbulence becomes more dominant. This change in airflow behavior reduces the cooling efficiency, explaining the decline in performance observed with configurations featuring a higher number of outlets. Consequently, the one-outlet configuration was selected for further simulations, as it achieved the best overall cooling performance across all speed regions.

With the 5-inlet and 1-outlet configurations established, the study evaluated the thermal performance of 7 battery pack arrangements under varying airflow conditions. The funnel configuration (54322) emerged as the most effective in the low-speed region, leveraging its narrow tail to accelerate airflow and enhance cooling. However, its performance deteriorates significantly in the high-speed region, where airflow instability undermines its heat dissipation capability. The rise in temperature in the high-speed region is attributed to the over-acceleration of tail airflow caused by the increasing inlet airflow speed. In the funnel configuration, this phenomenon leads to excessive heat accumulation at the end units of the 16-battery pack, as the high-velocity airflow reduces the residence time of air over the battery surfaces, impairing effective heat transfer. Consequently, the end units experience localized overheating, compromising the overall thermal management efficiency of the configuration in the high-speed region.

Among all configurations, the trapezoidal (wide-base) configuration (4543) exhibited the most balanced performance across all speed regions. In the low-speed region, its cooling performance was comparable to the funnel configuration, with only a marginally higher temperature. In the mid and high-speed regions, it maintained thermal stability and kept the battery temperature within the optimal operating range of 45 °C. This versatility makes the trapezoidal (wide-base) configuration a robust choice for drone applications, where low-speed cooling is critical during startup and hovering phases, yet adequate thermal management at higher speeds remains essential.

The results highlight the importance of carefully selecting inlet, outlet, and battery arrangement configurations to achieve optimal thermal management. For drone operations, where low-speed airflow dominates, the trapezoidal (wide-base) configuration is recommended for its consistent performance and ability to maintain battery temperatures within safe operational limits under varying airflow conditions.

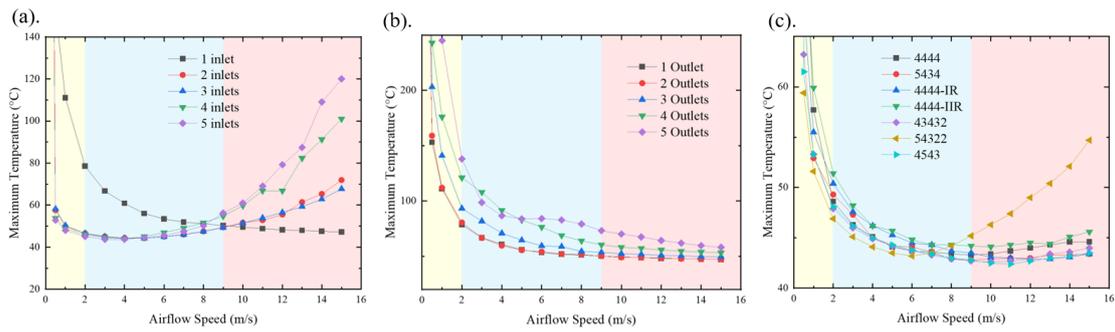

**Fig. 5** (a). Heat dissipation performance with varying inlet numbers. (b). Heat dissipation performance with varying outlet numbers. (c). heat dissipation performance of battery pack configurations.

## 4. Conclusions

The research conducted on the 16-battery pack provides valuable insights into optimizing thermal management through strategic design and simulation. The study systematically analyzed the influence of the inlet and outlet configurations, airflow conditions, and geometric arrangements on cooling performance. Results demonstrated the significance of balancing airflow dynamics to achieve efficient heat dissipation, particularly under low-speed conditions relevant to drone operations.

The key findings highlight that the 5-inlet and 1-outlet configuration offers superior cooling efficiency, maintaining low maximum temperatures across various airflow speeds. The trapezoidal (wide-base) configuration emerged as the most cooling-efficient arrangement, exhibiting balanced performance in low- and high-speed regions. This configuration is particularly suitable for drone applications, where thermal management is essential during critical stages like takeoff and hovering. While other configurations, such as the funnel and irregular triangular setups, showed strengths in specific related speed ranges, they lacked the versatility required for consistent performance.

The PCM integrated into the system demonstrated effective temperature regulation, with a phase change duration spanning approximately 12.5 min under different tested conditions. This case highlights its role in stabilizing thermal behavior and enhancing overall cooling performance.

In conclusion, the study provides a foundational framework for designing efficient thermal management systems for energy-dense battery packs. Future work will focus on experimental validation and extending the analysis to larger-scale battery systems or alternative thermal management techniques. The findings contribute to advancing cooling solutions for applications requiring compact and reliable energy storage, such as portable energy storage systems and other lightweight devices.

**Credit authorship contribution statement**

Xuguang Zhang: Resources, Writing – original draft, Writing – review & editing. Hexiang Zhang: Resources, Writing – review & editing. Amjad Almansour: Resources, Writing – review & editing. Mrityunjay Singh: Resources, Writing – review & editing. Hengling Zhu: Resources. Michael Halbig: Resources, Writing – review & editing. Yi Zheng: Resources, Writing – review & editing, Project administration, Supervision.

**Declaration of competing interest**

The authors declare that they have no known competing financial interests or personal relationships that could have appeared to influence the work reported in this paper.

**Data availability**

Data will be made available on request.

**Acknowledgments**

This project is partially supported by the National Aeronautics and Space Administration Glenn Research Center Faculty Fellowship and the National Science Foundation through grant number CBET-1941743.

**Nomenclature**

| | | | |
|---|---|---|---|
| $T$ | Temperature (°C) | $r_{os,i}$ | Outer sheath inside radius (mm) |
| $q_{gen}$ | Surface heat flux (W/m²) | $r_{os,o}$ | Outer sheath outside radius (mm) |
| $T_{ideal}$ | Optimal operating temperature (°C) | PCM | Phase change material |
| $T_{PCM}$ | PCM melting temperature (°C) | $k_{PCM}$ | Thermal conductivity of PCM (W/m·K) |
| $\Delta H$ | Latent heat (J/kg) | $V$ | Inlet airflow velocity (m/s) |
| $c_p$ | Specific heat (J/kg·K) | $r_{is,i}$ | Inner sheath inside radius (mm) |
| $\rho_{PCM}$ | Density of PCM (kg/m³) | $r_{is,o}$ | Inner sheath outside radius (mm) |
| $k_s$ | Sheath thermal conductivity (W/m·K) | $\rho_s$ | Sheath density (kg/m³) |
| $c_{p_s}$ | Sheath specific heat (J/kg·K) | $T_i$ | Initial temperature (°C) |